# Enhancing Trust in eAssessment - the TeSLA System Solution

Malinka Ivanova[1], Sushil Bhattacharjee[2], Sébastien Marcel[2], Anna Rozeva[1], Mariana Durcheva[1]

[1]Technical University of Sofia, 8 Kl. Ohridski blvd., Sofia, Bulgaria
`{m_ivanova, arozeva, m_durcheva}@tu-sofia.bg`
[2]Idiap Research Institute. Rue Marconi 19. CH-1920 Martigny, Switzerland
`{sushil.bhattacharjee, sebastien.marcel}@idiap.ch`

**Abstract.** Trust in eAssessment is an important factor for improving the quality of online-education. A comprehensive model for trust based authentication for eAssessment is being developed and tested within the scope of the EU H2020 project TeSLA. The use of biometric verification technologies to authenticate the identity and authorship claims of individual students in online-education scenarios is a significant component of TeSLA. Technical University of Sofia Bulgaria (TUS), a member of the TeSLA consortium, participates in large-scale pilot tests of the TeSLA system. The results of questionnaires to students and teachers involved in the TUS pilot tests are analyzed and summarized in this work. We also describe the TeSLA authentication and fraud-detection instruments and their role for enhancing trust in eAssessment.

## 1    Introduction

Evaluation of acquired knowledge and skills in the educational process is the final phase of the teaching and learning triad. The learning outcomes and competences of students are evaluated through performance of assessment tasks and the results obtained could be used as a valuable feedback for improving the teaching and learning. E-assessment has turned out to attract research interest and efforts due to the enhancement of online and distant education. A variety of tools and software solutions provide the technological infrastructure and architecture for supporting and facilitating the eAssessment. The main challenge to broad adoption of eAssessment by the educational authorities is the guarantee of trust: (1) trust in the identity of the examinee and (2) trust in the authorship of submissions. This issue of trust in eAssessment is the focus of a system that has been developed in the scope of the TeSLA[1] project, funded by the European Commission within the H2020 programme. TeSLA aims to provide trust-based authentication and authorship analysis for an eAssessment process. In TeSLA the core functionality for authenticating the identity of a student is provided by several biometric and forensic *instruments*:

---

1    tesla-project.eu

- biometric authentication instruments for face recognition (FR), and voice recognition (VR),
- authorship verification instruments for forensic analysis (FA) and plagiarism detection (PD), as well as
- security instruments for face anti-spoofing (FRA), voice anti-spoofing (VRA), and certificate-authorization (CA).

Some of these instruments are discussed in later sections of this paper.

The main goal of TeSLA, however, is to propose, develop and test a general framework for e-authentication using such instruments. The framework has been designed to easily integrate other instruments (implementing different authentication methods) or instruments similar in functionality to the existing ones but by different providers. The TeSLA framework, combined with such instruments, proposes a flexible, adaptable and trustworthy assessment environment to students and teachers in online and blended educational institutions [1].

- Flexibility: TeSLA enables the student to choose the time and geographical location to carry out the assessment tasks.
- Adaptability: the TeSLA system can be configured to suit different assessment scenarios, educational contexts, and varying legal requirements, including its capacity for eAssessment of students with special educational needs and disabilities [2].

One important aspect for the successful adoption of the TeSLA system for eAssessment is its trustworthiness. Peytcheva-Forsyth [3] discussed the importance of trust in eAssessment: "The lack of trust in e-assessment could ruin the public support that higher education system depends so heavily upon." She points out the technological solution proposed by TeSLA for student authentication and authorship verification as a means of building trust between students and educational institutions. Challenges related to the potential increase of cheating in eAssessment as well as for plagiarism were discussed in [4] together with proposals of students and teachers for their prevention and reduction. Mellar et al. [5] studied the trustworthiness of eAssessment in respect to cheating. They provided insights into the possibilities of authentication and authorship verification technologies for effectively addressing this issue, and pointed out the need for using a variety of approaches in conjunction with these technologies, to address the issue of cheating effectively. Saad et al. [6] discussed the American point-of-view about the quality of online education compared to the traditional classroom They reported that 49% of Americans believe that the employers do not view online education in a positive way, and 45% of Americans do not trust that online testing and grading are objective. The authors concluded that Americans see the benefits and power of online education, but in order to be competitive with traditional education, it must demonstrate high standards in instructional design, testing and grading. A report by the European Commission [7] examined new models of learning and teaching in higher education and shows that its quality is the basis for trust in certification recognition. In the context of online education the report suggests the need for trustworthy mechanisms for identity-verification of students, new forms for authentication, and new market for assessment services. Clearly, trust in eAssessment is a key point for the recognition of students' learning outcomes, their grading and certification.

Our paper focuses on investigation and analysis of trust issues related to the TeSLA system as observed during the large-scale pilot conducted at the Technical University of Sofia (TUS). The main goal of the pilot was to test the TeSLA framework for identity- and authorship-verification. The research questions posed in this work are:
- What are the factors determining trust in eAssessment?
- How can trust in eAssessment be achieved?
- How could the TeSLA system enhance trust in eAssessment?

The main contributions of this paper are the following:
1. we provide an analysis of the questionnaires and surveys concerning trust, filled in by participants of the pilot at TUS;
2. we present the trust-model implemented in the TeSLA system; and
3. we describe the role and functionality of the TeSLA instruments for authentication and fraud-detection as an important factor in enhancing trust in eAssessment.

The paper is structured as follows. Section 2 contains a review of trust models implemented in eAssessment. Section 3 presents an analysis of surveys obtained from students and teachers involved in the pilot at TUS concerning different aspects of trust in eAssessment. Section 4 highlights the technical aspect of the trust e-authentication model implemented in the TeSLA system by analyzing the biometric and anti-spoofing instruments, and Section 5 concludes the paper by summarizing the findings and results.

## 2  Trust models in eAssessment - state-of-the-art

Several trust models in eLearning and eAssessment have been proposed by researchers and academics, which outline the basic factors influencing trust. A review of some significant works in this area will provide better understanding of the challenges related to trust.

Wang [8] pointed out the need for building and maintaining trust among students in eLearning courses. The author suggested deeper exploration of the factors influencing trust and proposed a socio-technical framework with 12 factors classified in four groups:
- credibility – the previous reputation of the eLearning system and the instructor,
- design – of informative and graphical components of the eLearning system,
- instructor socio-communicative style – instructor's behavior and manner of communication, and
- privacy, and security of the eLearning system.

Wongse-ek et al. [9] considered the insufficient trust in learning activity as a cause for students' anxiety. The authors defined trust as "the belief and confidence of a student" in teaching agents. They proposed a conceptual model for a student's trust based on three groups of factors: (1) the student's assumption to trust, (2) the student's perception about the trustworthiness of the teaching activities, and (3) the pedagogical context concerning the student's readiness for a given teaching activity.

Kiennert et al. [10] investigated issues concerning trust of students and educators in eAssessment with a focus on providing mechanisms for security and protection of private data in the TeSLA system. The adopted solution was based on implementation of certificate-authorization and public key infrastructures [11].

Okada et al. [12] presented a model for trust-based e-authentication system taking into account the TeSLA system features, leading to trust among students, teaching staff, technical staff and quality assurers. The authors elaborated several features of the TeSLA system that provide for trust in eAssessment, which:
- should not affect students' and teachers' performance,
- should not impact students' experience adversly,
- should ensure a fair assessment process,
- should not fail or be compromised, and
- keeps user-data private and secure.

Marsh [13] analyzed trust as a computational concept. The author considered three levels of trust, namely:
- *Basic trust* - that is, the general trusting disposition of an agent A at time T,
- *General trust* - represents the trust that agent A has on agent B at time T and
- *Situational trust* - the amount of trust that one agent A has in another, taking into account a specific situation.

Miguel et al. [14] offered a normalized trustworthiness model. The authors reported that "from the results comparing manual evaluation and trustworthiness levels, it can be inferred that it is viable to enhance security in e-assessment by modeling and normalizing trustworthiness behaviors".

The connection and interrelationship of trust in eAssessment and the educational quality has also been explored [15]. The research implements a framework with standards for achieving quality and building trust in online assessment developed within the scope of the TeSLA project. Its elements include: policies, structures, processes and resources for quality assurance in eAssessment; learning assessment; security of eAssessment system; infrastructure and resources; students and teachers support; learning analytics; public information.

Analysis of the literature shows that trust could be explored from several aspects. In some cases it is connected to the competences of teachers, their teaching and communication style, as well as their domain subject expertise. In other cases, it is related to the features and functionality of the eLearning system concerning security and data privacy issues. Another part of the models propose complex solutions for building and maintaining trust, which concern the influence of social, technical, socio-technical, cultural and other factors that affect the effectiveness and quality of education. In the context of the TeSLA project, several factors, classified in three groups, are considered for the achievement of a trustworthy eAssessment process:
- *Behavioral* - concerns the performance of all participants in eAssessment, i.e., examinee, examiners, technical, administrative and legal staff;
- *Socio-cultural* - trust of external universities and organizations in the fairness of the eAssessment process and the accuracy of the achieved and assessed outcomes of students;

- *Technical* - trust of all participants in the features and functionality of the system for eAssessment.

This work focuses on the technical aspect of trust concerning eAssessment.

## 3 Trust in eAssessment - analysis of students' and teachers' surveys during the large-scale TeSLA pilot

TUS is one of the seven TeSLA partner universities running pilot programmes (referred to simply as pilots from hereafter) to test the eAuthentication functionalities of TeSLA. Findings on student experiences with the system have been reported in [16] and [17]. In the course of the last large-scale pilot, 452 students and 8 teachers, from several faculties of TUS, completed the post-questionnaire. The results presented summarize their opinions about trust in online assessment and in eAssessment supported by the TeSLA system after finishing the pilot experiment. To understand the obtained results, it is important to state that the assessment activities and examinations in TUS are generally performed face-to-face, along with certain forms of online assessment. The survey-questions are classified in six groups:

- *Trust in online assessment system:*

The answers of surveyed students and teachers regarding their trust in online assessment system show that about 22% of the students (5.3% vote with "strongly disagree" and 16.2% with "disagree") do not accept the idea of online assessment at all, whereas there are no such opinions among the teachers. Interesting fact is that about 37.4% of the students and half of the teachers are neutral regarding trust in online assessment. The reason that many students and teachers have no opinion can be explained by the fact that TUS is a blended learning university and they do not have much experience in eAssessment. Nevertheless, 41% of the students (30% of them voted "agree" and 11% "strongly agree") and almost all of the teachers trust the online assessment. So far as the raising of trust in online assessment by using the TeSLA system is concerned 38.3% of the students reply with "agree" and 7.9% of them with "strongly agree". A significant portion (37.9%) of the students are neutral. This result could be explained with the fact that they completed very few eAssessment activities (many such students performed only one assessment activity), and therefore, could not become very familiar with the advantages of the system. Only part of students tested all TeSLA instruments in two or three assessment activities. In contrast, the majority of the teachers became aware of the benefits of the TeSLA system and therefore they found eAssessment with this system trustworthier than the students did.

- *Cheating in online assessments:*

The statement "It is easy to cheat in online assessments" received quite polarized responses from students. Whereas 27.6% of the students think that it is *not* easy to cheat during online assessment, 41% of them are neutral and 31.4% agree that it is easy to cheat. The teachers' answers show that only few of them confirm the possibility for cheating in online assessment. These answers could be explained with the fact that cheating is one of the main problems both for face-to-face and online assessment.

- *Trust in the prevention of cheating by e-authentication in eAssessment:*

  From the students' and teachers' responses to the statement "The use of e-authentication (security measures) for online assessment will make it more difficult for students to cheat" it can be observed that teachers are absolutely convinced of the benefits of the security measures for online assessment. This opinion is shared by 33.3% of the students, whereas 42.2% of them are neutral and 24.4% do not consider that e-authentication is an effective measure against cheating. In our opinion, the reason for this is insufficient or incorrect understanding of the biometric identity-verification methods among some students.

- *Trust in the obtained assessment results:*

  Responses to the statement "E-authentication will help me trust the outcomes of the online assessment" show that both surveyed groups of students and teachers are convinced of the usefulness of e-authentication for online assessment (40% of students and most of the teachers' vote with "agree" and "strongly agree") and 44.5% of students are neutral. Nevertheless, there are students (15.5%) who do not trust the results of eAssessment. This is understandable and can be considered normal.

- *Trust in ensured security and data privacy:*

  The answers to the statement "Using the TeSLA system increased my confidence in the security of my students' assessment" confirm the positive attitude of almost all of the teachers and students (33%) regarding the security measures implemented in the TeSLA system and their effectiveness during the online examination. A majority of the students (51.8%) are neutral regarding the security issues, due to insufficient understanding of the security mechanisms in the TeSLA system.

  The statement concerning data privacy (a question for students): "While I used the TeSLA system, I was confident that my personal data was being treated properly", 44% of students trust the TeSLA system and mention that they are not worried about their personal data. Less than 18% are not confident that their personal data are being treated properly. This is a very good achievement within the TeSLA pilot.

- *Trust in the TeSLA e-authentication instruments:*

  Half of the students expressed their trust in the tested TeSLA e-authentication instruments during the pilot (43.3% of them vote with "agree" and 6.7% with "strongly agree") and only 14% voted negatively.

  The students at TUS have experience mainly with face-to-face examinations and just a small part of the assessment activities is performed online. Also, 62% of the surveyed students declare that they do not have any previous experience with online assessment. So, it can be seen from students' answers to this statement that their trust in the tested TeSLA instruments is very high. The trust in every single TeSLA instrument is presented in Table 1.

**Table 1.** Students' trust in TeSLA instruments. (The numbers represent percentages).

| TeSLA instrument | | FR | VR | KSD | FA | PD |
|---|---|---|---|---|---|---|
| Number of students tested the instrument | | 60.4 | 31.9 | 22.8 | 13.3 | 13.1 |
| "If the instrument was used in my future online assessments, it would increase my trust in the assessment result". | Strongly disagree | 9.9 | 15.3 | 10.8 | 5.1 | 5.1 |
| | Disagree | 16.5 | 16.7 | 14.7 | 6.8 | 10.2 |
| | Neutral | 38.2 | 45.1 | 45.1 | 52.5 | 39 |
| | Agree | 28.3 | 19.4 | 26.5 | 30.5 | 44.1 |
| | Strongly agree | 7 | 3.5 | 2.9 | 5.1 | 1.7 |
| "I would be willing to use the instrument in all my future assessments." | Strongly disagree | 15.8 | 22.2 | 13.7 | 6.8 | 6.9 |
| | Disagree | 14.3 | 18.1 | 16.7 | 11.9 | 20.7 |
| | Neutral | 37.9 | 40.3 | 40.2 | 45.8 | 39.7 |
| | Agree | 24.3 | 14.6 | 26.5 | 30.5 | 27.6 |
| | Strongly agree | 7.7 | 4.9 | 2.9 | 5.1 | 5.2 |

A significant number of the surveyed students are neutral about whether the use of TeSLA instruments in their future online assessment activities would increase trust in their assessment results. At this stage of the pilot experiment they have very little experience with the TeSLA instruments and cannot evaluate their full power for online examination. Through their participation in the pilot students have become aware, to a certain extent, of the advantages of the e-authentication platform for performing online assessments. The students' vote is similar to the statement regarding their willingness to use the instruments in their future assessment activities.

- *Trust between students and the university:*

The statement under vote was: "I feel that my university does not trust its students." According to teachers' responses none of them believes that the university does not trust its students. The student vote shows that 39.4% of them do not share the opinion that the university uses the TeSLA system because it does not trust them. 37.9% of the students are neutral and 22.6% agree with this statement (Table 2).

- *Trust between students and teachers:*

Regarding the statement "eAssessment will increase the trust between teachers and students" no teacher doubts in that. However, 22.8% of the students are skeptic that online assessment could increase the trust between them and teachers, 47.9% of them does not have an opinion and 29.2% of them agree. That is an interesting phenomenon, and may be because in some aspects young people are more skeptical and believe that teachers will always have the last word, regardless of the form of assessment.

**Table 2.** Trust among students, teachers, university and external bodies (The numbers represent percentages).

| Statement | Scale | Students' vote | Teachers' vote |
|---|---|---|---|
| "I feel that my university does not trust its students" | Strongly disagree | 11.5 | 50 |
| | Disagree | 27.9 | 37.5 |
| | Neutral | 37.9 | 12.5 |
| | Agree | 18.2 | 0 |
| | Strongly agree | 4.4 | 0 |
| "eAssessment will increase the trust between teachers and students" | Strongly disagree | 5.1 | 0 |
| | Disagree | 17.7 | 0 |
| | Neutral | 47.9 | 12.5 |
| | Agree | 24.8 | 62.5 |
| | Strongly agree | 4.4 | 25 |
| "Usage of e-authentication in online assessment will increase the trust that other universities and employers will have with the outcomes of ONLINE assessments" | Strongly disagree | 4.9 | 0 |
| | Disagree | 11.1 | 0 |
| | Neutral | 46.8 | 12.5 |
| | Agree | 31.3 | 62.5 |
| | Strongly agree | 6 | 25 |

- *Trust between students, other universities and employers:*

The voted statement is: "Usage of e-authentication in online assessment will increase the trust that other universities and employers will have with the outcomes of online assessments". Regarding trust between students and other universities and employers, the results show that 37.3% of the students believe that online assessment will increase the trust in their achievements and perhaps this is due to the fact that the subjective factor (opinion of the teacher about the student) disappears. Teachers' opinion is that online assessment will undoubtedly increase the confidence in students' results.

The analysis of the post-questionnaire from different viewpoints referring to trust in TUS, highlights its importance for different parties (students and teachers) and support for e-authentication and prevention of cheating in the eAssessment process.

## 4   Trust model implemented in the TeSLA system

The review of trust based eAssessment shows that it is most strongly related to trustworthy e-authentication. Trustworthiness in TeSLA e-authentication and authorship verification system is provided by different classes of instruments. The trust model involving them is shown on Figure 1.

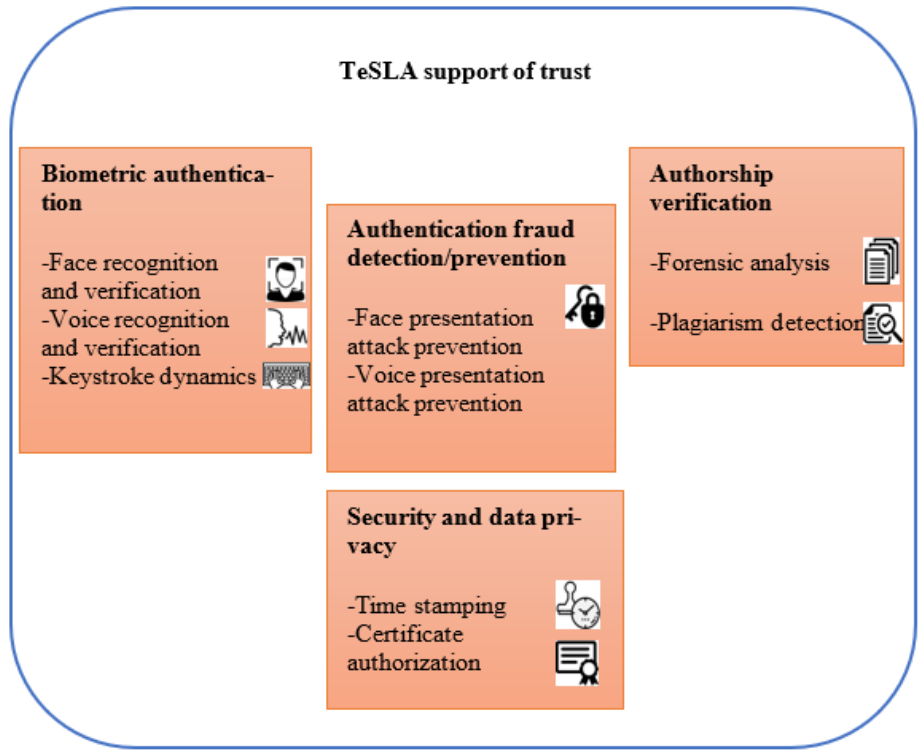

**Fig. 1.** Trust model in the TeSLA system from technical point of view

The current implementation of TeSLA relies mainly on two biometric authentication instruments, namely, face recognition (FR), and voice recognition (VR) and is further augmented by the soft-biometrics instrument keystroke dynamics (KD). TeSLA also includes specific instruments to detect spoof attacks, namely, the face-recognition anti-spoofing (FRA) instrument and the voice-recognition anti-spoofing (VRA) instrument providing the enhancement of the trust model implemented in TeSLA. Here, we discuss the biometric authentication instruments (Section 4.1) and anti-spoofing instruments (Section 4.2) separately.

### 4.1 Biometric Authentication Instruments in TeSLA

The biometric authentication instruments function in the same general fashion, that is, in two modes: enrollment and verification (also called probe in biometrics parlance). Before the identity of person A to be verified by a given instrument, the person A needs to be enrolled for that instrument. During the enrollment, the instrument uses appropriate biometric samples of A to generate a set of enrollment templates, which are stored in an appropriate database, indexed by the identity (A). During verification, the instrument takes as input a new biometric sample (termed a probe sample), as well as a claimed identity – A, in our example – generates a probe template from the given sample, and compares the probe template with previously enrolled templates for the claimed identity. If the match between the probe template and the enrollment templates is sufficiently strong, the claimed identity is accepted, that is, the presented probe sample is considered to belong to the person A. In this section we present brief descriptions of the TeSLA biometrics instruments listed earlier.

**Face Recognition** based on digital images has been a topic of intense research for several decades [18]. In recent years, deep learning [19] based approaches to FR have been shown to outperform previous methods by a wide margin. At present, FR methods based on deep convolutional networks (DCNN), such as the VGG-Face network [20] and FaceNet [21] are considered state of the art. The FR instrument in TeSLA is also based on a DCNN. In placed near the top of the leader-board in one of the largest industrial benchmark tests [2] for FR systems. For TeSLA, the DCNN has been retrained with GDPR compliant training data.

To enroll a student in the FR instrument, a short video (5 ~ 7 sec.) of the student is recorded for capturing frontal and lateral views of the student's face. Frames of this video are then used to generate an enrollment face-template for the student.

Subsequently, during a given course activity, short face video-clips of the student are periodically captured by TeSLA. TeSLA transmits these video clips, along with the student's claimed identity to the FR instrument, which processes video-frames in these clips, generates probe templates, and compares the probe templates to the enrollment templates corresponding to the claimed identity. The templates (enrollment and probe) are represented as real-valued vectors, and are compared using the Cosine similarity measure[3]. If the Cosine similarity between the probe template and enrollment template is higher than a certain preset threshold, the claimed identity of the student is considered to be valid for the corresponding probe video-frame.

Tests of the DCNN used here, based on non-TeSLA data (the Megaface[2] dataset) showed total error (false accept rate (FAR) + false reject rate (FRR)) of approximately 1.1%. (Ideally, both error-rates should be 0, but the ideal case is rarely achieved in practice.)

**Voice Recognition** often also referred to as automatic speaker recognition (ASR), attempts to characterize personal voice-patterns and to verify the identity of a person based solely on the person's voice and speech patterns. Reynolds [22] provides a nice overview of basic VR techniques. The VR method used in TeSLA's VR instrument uses i-Vectors [23, 24] for modeling speaker representations. I-Vector based approaches constitute state-of-the-art VR technologies, especially for short utterances.

For enrollment, several samples, each approximately 15 seconds long are required for each student. Typically, it is required students to provide 15 voice-samples, recorded over three sessions (five samples per session). Ideally there should be an interval of several days between consecutive sessions. The VR instrument generates and stores an i-Vector representation (a 400-element vector of real-valued numbers in the current implementation) for each enrollment voice-sample of a given student. The enrollment i-Vectors are, of course, indexed by the TeSLA-identifier of the student in question. It is very important to use multiple samples for enrollment, as this allows the instrument to model the natural inter-session variability in human voice much more accurately. For the same reason, it is also important to collect enrollment samples over multiple sessions, at different times, rather than collecting all enrollment samples in a single session.

---

[2] http://megaface.cs.washington.edu/results/facescrub.html

[3] https://en.wikipedia.org/wiki/Cosine_similarity

During a subsequent course activity, TeSLA periodically captures a voice-sample of the student and passes it on to the VR instrument, along with the claimed identity of the student. The instrument produces an i-Vector representation of the input voice-sample, and compares it to all the enrolled i-Vectors for the claimed identity. The Cosine similarity measure is used for comparing i-Vectors. The sample is said to match the claimed identity if the probe i-Vector matches any one of the enrollment i-Vectors for that identity to a sufficient degree, that is, if the Cosine similarity is above predetermined threshold for at least one enrollment i-Vector. Evaluations with non-TeSLA datasets show that this method can achieve FAR = 8.85% and FRR of 8.31% [33]. In other words, statistically speaking, in one out of every 11 or 12 attempts (~ 8.85%), a student **A** may succeed in impersonating another student, **B**, if only VR was used to verify the identity of students. Similarly, in statistical terms, for one out of 12 voice-samples, the VR instrument may determine, wrongly, that the sample does not come from the student with the claimed-identity. Either way, these expected error-rates are reasonably low, and give no cause for alarm. In future, when the VR instrument in TeSLA will be tuned for optimal performance using data collected during these large-scale pilots, we expect the corresponding error rates to be significantly lower. Moreover, TeSLA authentication relies on multiple biometric instruments, which leads to significantly lower error-rates than when using only a single instrument.

**Keystroke Dynamics** instrument attempts to verify the claimed identity of a student by analyzing the computer-keyboard typing style of the student. As the student types on the keyboard, TeSLA collects two pieces of information about him/her: dwell time, that is the time that a key remains depressed, and flight time – the time interval between two consecutive keystrokes.

During enrollment, the student is required to type in a 2500-word text. Statistics of the dwell-time and flight-time values collected during this session are used by TeSLA KD instrument to create and store a model of the student's typing pattern. During a course activity that involves typing on the same keyboard, the dwell-times and flight-times collected intermittently by TeSLA are compared with the enrollment model, to verify if the typing pattern of the person performing the course activity is statistically conformant with the typing pattern of the claimed identity. Preliminary tests with reduced training data (150 keystrokes instead of 750) showed FAR and FRR both to be close to 2%.

### 4.2 Enhancement of trustworthiness in e-authentication by detecting and preventing cheating attacks

**Presentation Attack Detection (PAD)** also called anti-spoofing, is one of the strongest threats to biometrics-based identity verification. Figure 2 (inspired by [27]) shows the possible points of attack of a generic biometrics authentication system. Attacks on the biometric sensor using fake biometric samples (position 1) are termed 'Presentation Attacks'. Among the various points of attack, the biometric sensor (camera, microphone, fingerprint-sensor, and so on) is the most vulnerable, as this is the component of the biometric system that is most exposed to the outside world. Attacks performed on the biometric sensor are called spoof-attacks, or more formally, presentation attacks (PA). Correspondingly, countermeasures for detecting PAs are referred to as presentation attack detection (PAD) methods. The past two decades have seen intensive research on developing PAD methods for various biometrics modalities [28].

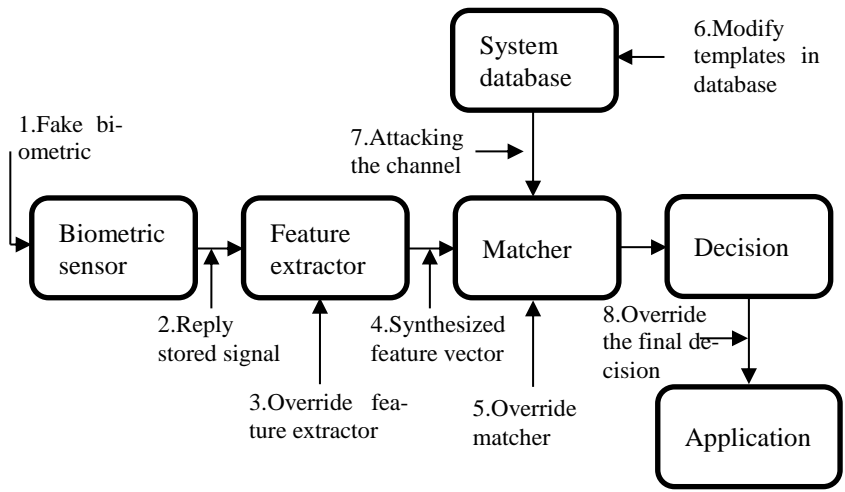

**Fig. 2.** Possible points of attack on a generic biometric system (figure inspired by [27]).

In a PA, an attacker, **A**, attempts to spoof the identity of another person, **B**, by presenting to the biometric system a biometric sample corresponding to **B**. For example, **A** may impersonate **B** by presenting a printed photograph of **B** to the camera of a FR system. The task of a PAD system is to detect whether the presentation to the biometric sensor is *bona fide* or is it a PA. TeSLA includes two PAD instruments, namely, the face-recognition anti-spoofing (FRA) instrument and the voice-recognition anti-spoofing (VRA) instrument. The two instruments are briefly described in the following sections.

*Face Presentation Attack Detection.* PAs on a FR system may be performed in several ways [28]:
- Print attack: the attacker **A** uses a printed facial photograph of an enrolled subject **B** as a PA.

- Replay attack: **A** presents to the camera a digital photograph or video of **B** played back on an electronic screen (e.g. a smartphone or tablet-device).
- Mask attack: **A** may present to the camera a custom-made mask (rigid or flexible) corresponding to the face of **B**.

The first two kinds of attacks are two-dimensional (2D) in nature, whereas masks are 3D. In the FRA instrument of TeSLA we consider only 2D attacks, that is, print attacks and replay-attacks. In TeSLA, we do not consider custom-made mask attacks for two reasons:

- Manufacturing highly realistic custom-made masks is a specialized and expensive process, and we do not expect students to go to such lengths to cheat in an exam;
- Detection of mask-based PAs is best performed using specialized hardware – typically involving infrared imaging [29]. In typical learning environments addressed by TeSLA, such as where a student performs course activities using a laptop of desktop computer, we cannot expect such specialized hardware to be routinely available.

Numerous PAD approaches can be found in the relevant scientific literature [28]. One popular approach for 2D face-PAD is based on image-quality measures (IQM). The idea behind the use of IQMs is the following: since 2D face-PAs are essentially photographs being re-captured by the biometric sensor (camera), it is reasonable to expect the image-quality of a PA to be perceptibly lower than that of a *bona fide* presentation. Several works [30, 31] have demonstrated the efficacy of various IQMs in detecting different kinds of 2D attacks.

The FRA instrument in TeSLA processes the same data that is provided as input to the FR instrument. This instrument is agnostic to the student's identity – its purpose is solely to determine whether a given input image is *bona fide* or a PA. It uses a set of 18 IQMs [30, 32] computed over each frame of face-video. A pre-trained two class classifier – specifically, a support vector machine (SVM) – classifies the 18-element feature-vector to detect presentation attacks. Tests on non-TeSLA datasets have shown that this approach achieves accuracy measured as the average classification error rate (ACER) – of approximately 5% to 10%, depending on the dataset [32].

*Voice Presentation Attack Detection.* Voice PAs are PAs performed on voice-based identity-authentication systems. Voice PAs may also be performed in several ways [28]:

- Replay attack: an attacker **A** records **B** speaking, say on a smartphone, and replays the recorded speech of **B** to the microphone that functions as the biometric sensor for the VR instrument.
- Voice conversion: the attacker **A** uses voice-conversion software to make previously recorded speech (say, recorded speech of A) to sound as if it has been spoken by **B**. The resulting converted speech may be played back to the microphone of the VR system.
- Voice Synthesis: the attacker **A** uses voice-synthesis software to synthesize the speech of **B** from text.

In the context of TeSLA we consider voice-conversion and voice-synthesis attacks to be highly unlikely. Therefore, the Voice Recognition anti-spoofing (VRA) instrument in TeSLA is designed to detect mainly voice replay-attacks.

As for the face modality, voice-PAD has also been a topic of intense research in recent years. Most works in voice-PAD have addressed the problem using various frequency-domain approaches. The VRA instrument in TeSLA computes a set of Mel-Frequency Cepstral Coefficients (MFCC) [33] and their derivatives from the given input speech sample. This instrument processes the same voice-samples that are provided as input to the VR instrument. The VRA instrument uses a Gaussian Mixture Model (GMM) based OCC to determine whether the input voice-sample is *bona fide*. In preliminary tests, using non-TeSLA datasets, this approach has shown a ACER of 1.01% [34].

### 4.3 Enhanced trust model in the TeSLA system

Trust in the eAssessment process supported by TeSLA is enhanced by the implementation of two different instruments for cheating attacks detection: face-recognition anti-spoofing and voice-recognition anti-spoofing instruments. They are designed to detect possible authentication fraudperpetrated by students while performing the assessment activities. Figure 3 illustrates the role of anti-spoofing instruments FRA and VRA for building and maintaining trust when applying different online assessment and examination scenarios.

The FRA instrument can detect the fraud when a printed facial photograph, digital image or video for authentication in eAssessment activities is used instead of the student actually present in front of the computer camera. The VRA instrument can detect if a person uses previously recorded voice-samples of a registered student for authentication in a real assessment activity. Thus, the anti-spoofing instruments in TeSLA enhance trust in eAssessment, and the achieved results by students.

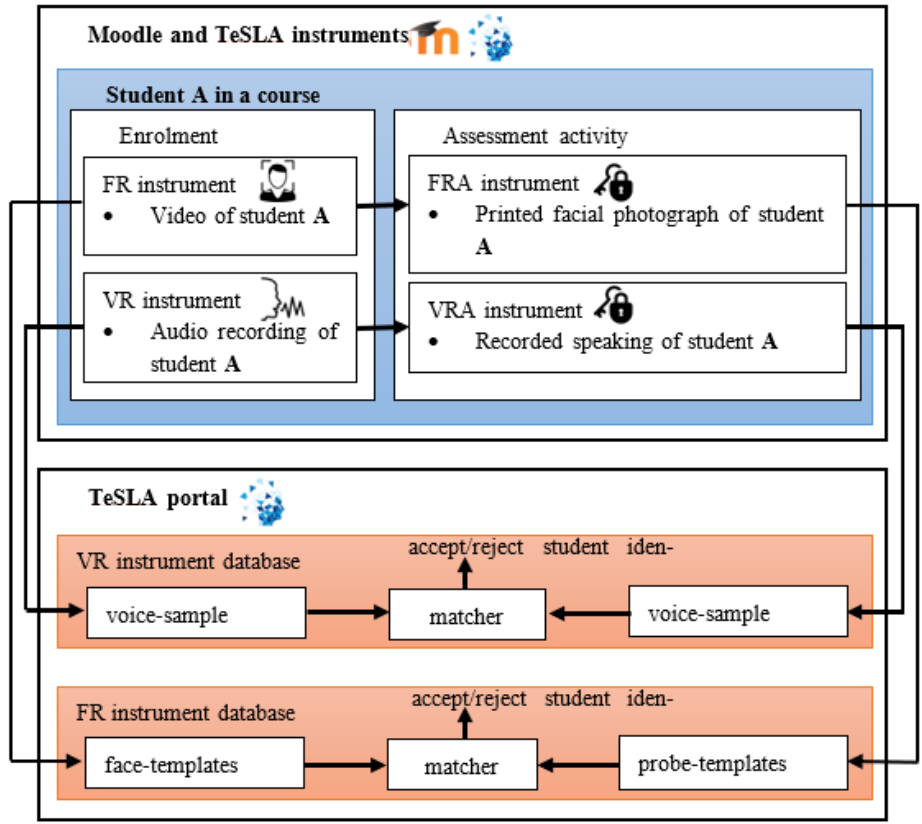

**Fig. 3.** Trust enhancement with anti-spoofing instruments FRA and VRA

## 5    Conclusion

The analysis provided in this paper shows that building and maintaining trust in eAssessment is a complex process and several factors need to be taken into account for ensuring trust. The literature review revealed different trust models implemented in eLearning and eAssessment practice. It turns out that a student's trust in online learning and assessment depends on three main factors: (a) the teacher's competence and expertise, (b) functionalities related to data security and privacy in the online learning environment, and (c) implemented communication model among all participants and stakeholders involved in the educational process. At the end of the large-scale pilot of the TeSLA system in TUS, students and teachers surveyed reported a relatively high-level of trust in the tested TeSLA instruments for biometric authentication, despite the limited experience they have in online examination.

    The students seem to be somewhat skeptical about the use of online assessment because of concerns regarding the ability of the system to detect fraudulent conduct during the authentication process. Teachers, on the other hand, expressed their acceptance of

online examination supported by a stable and trust-based e-authentication system. The anti-spoofing instruments (FRA and VRA) included in the TeSLA platform are designed to address concerns about the trustworthiness of the core biometric instruments (FR and VR, respectively). The use of such instruments can provide for enhanced trust model by detering cheating and fraudulent behavior in online assessment scenarios.

**Acknowledgements**

This work has been supported by the H2020 project TeSLA – An Adaptive Trust-based e-assessment System for Learning (grant №. 688520).